\documentclass[11pt]{article}
\usepackage[utf8]{inputenc}
\usepackage{times}

\usepackage[margin=1in]{geometry}
\usepackage{microtype}
\usepackage{amsmath,amsthm,amssymb}
\usepackage{graphicx}
\usepackage{booktabs}
\usepackage{array}
\usepackage{tabularx}
\usepackage{xcolor}
\usepackage{caption}
\usepackage{flafter}
\usepackage{float}
\usepackage{placeins}

\usepackage{tikz}
\usetikzlibrary{arrows.meta,positioning,shapes.geometric}

\definecolor{tlncol}{HTML}{1F6F8B}
\definecolor{ucncol}{HTML}{8B3A1F}
\definecolor{leakcol}{HTML}{B00020}
\definecolor{guardcol}{HTML}{1E7D32}
\definecolor{sealedcol}{HTML}{5B2C91}
\newcommand{\abadge}[3]{\node[font=\tiny\bfseries,fill=#1!14,text=#1,draw=#1,rounded corners=1pt,inner sep=1.4pt] at (#2) {#3};}
\newcommand{\stepbox}[4]{\node[draw=#4!70!black,fill=#4!8,rounded corners=2pt,align=left,text width=4.9cm,inner sep=4pt,font=\scriptsize] at (#1) (#2) {#3};}
\newcommand{\wirelbl}[4]{\draw[-{Stealth[length=2.4mm]},thick,#4] (#1) -- (#2) node[midway,font=\tiny,align=center,#4,fill=white,inner sep=1pt]{#3};}
\PassOptionsToPackage{hyphens}{url}
\usepackage[colorlinks=true,linkcolor=black,citecolor=blue!50!black,urlcolor=blue!50!black]{hyperref}

\newcommand{\tln}{\textsc{tln}}
\newcommand{\ucn}{\textsc{ucn}}
\newcommand{\pp}{pp}

\newcommand{\artifacturl}{\url{https://github.com/setloop-io/Privacy-Failure-in-Split-LLM-Training-The-Returned-Gradient-Nullifies-the-Decoys}}

\title{\textbf{Privacy Failure in Split-LLM Training, The Returned
Gradient Nullifies the Decoys}}
\author{Georgios Politis\\ {\small\texttt{geo.politis@gmail.com}} \and Setloop.io \and
  Evangelos Pappas\\ {\small\texttt{evanz.pappas@gmail.com}}}
\date{}

\begin{document}
\maketitle

\begin{abstract}
We present a systems-security case study of a two-node split-LLM training
system whose privacy evaluation passed while leaving an observable channel
untested. The Trusted Local Node (\tln) sends protected activations to the
Untrusted Cloud Node (\ucn), the \ucn{} returns its output, and \tln{}, holding
the private loss, returns the output gradient. The frame the \ucn{} receives mixes real rows
with decoys, and the loss ignores the decoys. Their gradients are exactly zero,
so the pattern of zeros reveals which rows were real.

We measure it with a protocol fixed in advance: a leak injected at known
strength to prove the instrument can see one, a shuffled-label control to prove
it does not report absent leaks, and a threshold set before the runs. Across
nine seeds, the zeros identified the real rows on every frame, 4,096 of 4,096
per run. An attack on the frame contents recovered about one extra token
per hundred over a constant-guess baseline ($+0.65$ to $+1.50$ percentage
points); the shuffled controls recovered nothing.

A second set of runs repeated this on a configuration that keeps model quality
within budget, so the finding is not confined to a setting nobody would deploy.
On both datasets, every such run passed the forward-channel privacy check and
the quality check, yet failed that same check once the returned gradient was
included. Clipping and noising each row of the gradient closed the leak for
about $0.01$ nats of held-out cross-entropy. The system is not thereby safe:
five classes of attack, including those accumulating observations across
training steps, were never measured.
\end{abstract}

\section{Introduction}

Split learning~\cite{vepakomma2018split} lets a data owner rent cloud compute for training without sending
raw examples: the trusted side sends activations and, because it alone holds the
private loss, returns the output gradients the untrusted side needs to train.
The privacy claim of such a system is that the cloud cannot read the
training data. The claim rests on an evaluation, and the evaluation rests on an
instrument.

The instrument examined here failed silently. The original evaluation of this
two-node split-LLM system instrumented the forward wire, the activations sent
to the cloud, and passed its privacy gate. The backward wire, which carries the
output gradient the trusted side returns to the cloud, was never declared a privacy
surface. In this implementation, excluding the decoy rows from the private loss
makes their returned gradients identically zero; that zero-support construction
is an implementation and system-design defect. The false assurance has a
distinct, more general evaluation cause: the observable gradient channel was
outside the declared adversary view, so the gate never tested it. This paper is
therefore a systems-security case study with a calibrated protocol applied to
one system, not a general methodology paper. Its contribution is a verified
diagnosis of that case and a channel-explicit evaluation, positioned against
three recent split-LLM evaluations, attack and defence alike
(Section~\ref{sec:related}).

Separating the real rows from the decoys matters because the decoys exist
precisely to hide which rows carry the private loss. Exact gradient support lets the
cloud discard all 48 decoys and isolate the 32 loss-bearing rows before any
content attack, collapsing the anonymity set that padding was meant to create.
This is a structural metadata disclosure, not text reconstruction; the content
evidence is separately bounded to the implemented frequent-token probe.

Section~\ref{sec:system} defines the system and its declared threat surface.
Section~\ref{sec:protocol} lays out the evaluation protocol, and
Section~\ref{sec:calibration} calibrates the instrument against injected leaks.
Section~\ref{sec:results} reports the structural disclosure, its replication,
and the mitigation runs; Section~\ref{sec:shape} reports how the signal
converts to a token advantage with depth, width, and budget.
Section~\ref{sec:related} audits three external evaluations and places the work
against the DP-auditing and split-LLM literature.
Section~\ref{sec:scope} records scope, limitations, and the verification trail.

\section{The system and the threat model}
\label{sec:system}

Two cooperating nodes train one LLM. We refer to them throughout by their
roles: the Trusted Local Node (\tln) holds the data and the
embedding head, and the Untrusted Cloud Node (\ucn) holds a middle stack of
transformer layers. For each training frame, \tln{} sends a protected latent
(the \emph{forward wire}); \ucn{} trains its stack and returns its output;
\tln{}, which alone holds the private loss, scores that loss and sends the
output gradient \emph{back} to \ucn{} (the \emph{backward wire}). Both tensors
are held by the untrusted node in the clear.

The defence on the wire is a latent-space bottleneck at width $D{=}64$, a
per-request rotation and permutation, decoy rows mixed among the real ones,
and boundary clipping and noise. Each 80-row frame carries 32 real rows and 48
decoys. The implementation and the artefacts call these decoys \emph{chaff},
after the radar countermeasure, and we use the two words interchangeably. The original evaluation declared the forward wire as the privacy
surface, instrumented it, and passed its gate. The backward wire was outside the
declared adversary view.

Table~\ref{tab:setup} sets out the evaluated configuration together with the
replication and mitigation runs.

\begin{table}[H]
\centering\scriptsize
\renewcommand{\arraystretch}{1.12}
\begin{tabularx}{\textwidth}{@{}>{\raggedright\arraybackslash}p{2.45cm}>{\raggedright\arraybackslash}X@{}}
\toprule
field & evaluated setup \\
\midrule
model and runtime & Qwen3-0.6B; Transformers 5.13.0 runtime; exact weights
SHA-256
\texttt{f47f7117\allowbreak 7f32bcd1\allowbreak 01b7573e\allowbreak c9171e6a\allowbreak 57f4f4d3\allowbreak 1148d38e\allowbreak 382306f4\allowbreak 2996874b} \\
topology & Main configuration and replication runs: split after layer 14,
resume at 26 (11 delegated layers). Mitigation runs: split 21/26 (4 delegated
layers). \\
corpus and implemented split & WikiText-2; corpus SHA-256
\texttt{78b6bfb9\allowbreak 0cfd718f\allowbreak 0c27d42b\allowbreak 1fd2231b\allowbreak 139d1dda\allowbreak 75d7d796\allowbreak e6a603b2\allowbreak e5cd7efe}.
The runner tokenizes the flat corpus into disjoint,
sequential fixed-width blocks, then slices train followed by evaluation; this
is not a document-level split. \\
frame and latent & 32 real $+$ 48 decoy $=$ 80 rows per frame; $D{=}64$. \\
boundary protection & Forward and returned activations: per-row $C{=}1.0$ and
Gaussian $\sigma{=}0.35C$. Main configuration and replication runs, outbound
gradient: open. Mitigation outbound gradient: $C{=}0.01$ and Gaussian
$\sigma{=}0.35C$. \\
known-plaintext latent probe & \texttt{coordinate\_plus\_invariants},
\texttt{invariant\_only}, and \texttt{invariant\_graph}; 20 epochs; batch 512; 3
restarts. \\
captured frames per run & Main configuration and replication runs: 4,096 $=$
2,048 probe-train $+$ 2,048 evaluation. Mitigation runs: 1,024 $=$ 512
probe-train $+$ 512 evaluation. \\
seed groups & 42--47 exploratory; 48--50 replication; 51--53 mitigation. \\
decision thresholds & Privacy: gate statistic $> +1.0$~\pp{} fails. Utility:
held-out cross-entropy increase $\le 0.35$~nats passes. \\
inference & Inferential unit: one independent training seed. Within-run
uncertainty is frame-clustered. \\
\bottomrule
\end{tabularx}

\caption{Experimental setup for the main configuration and for the replication
and mitigation runs. Full fingerprints identify the exact model
weights and corpus used by the evaluated runner.}
\label{tab:setup}
\end{table}

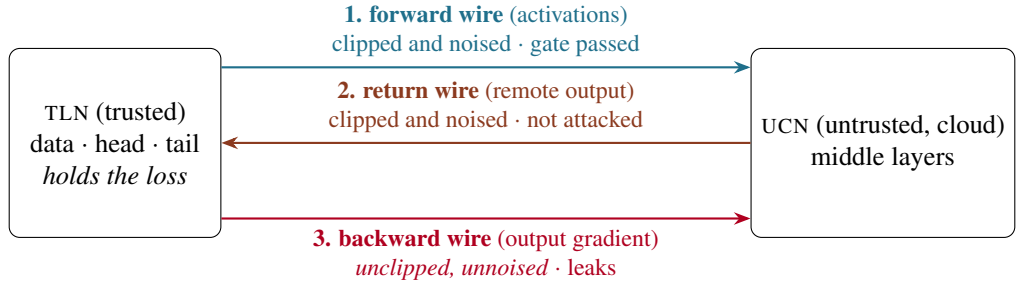
\begin{figure}[H]
\centering
\begin{tikzpicture}[
  node/.style={rectangle,rounded corners,draw,minimum height=2.5cm,minimum width=2.8cm,align=center,font=\small},
  chan/.style={-{Stealth[length=2.2mm]},thick},
  lbl/.style={font=\footnotesize,align=center}
]
\node[node] (t) {\tln{} (trusted)\\data $\cdot$ head $\cdot$ tail\\\emph{holds the loss}};
\node[node,right=7.0cm of t] (k) {\ucn{} (untrusted, cloud)\\middle layers};
\draw[chan,tlncol] ([yshift=10mm]t.east) -- ([yshift=10mm]k.west)
  node[lbl,midway,above]{\textbf{1. forward wire} (activations)\\clipped and noised $\cdot$ gate passed};
\draw[chan,ucncol] (k.west) -- (t.east)
  node[lbl,midway,above]{\textbf{2. return wire} (remote output)\\clipped and noised $\cdot$ not attacked};
\draw[chan,leakcol] ([yshift=-10mm]t.east) -- ([yshift=-10mm]k.west)
  node[lbl,midway,below]{\textbf{3. backward wire} (output gradient)\\\emph{unclipped, unnoised} $\cdot$ leaks};
\end{tikzpicture}
\caption{The three hops of one training step. Only \tln{} can compute the
gradient, because only \tln{} holds the loss, so hop~3 travels left to right
just as hop~1 does: ``backward'' names the pass the tensor belongs to, not the
direction it moves. Hops~1 and~2 are clipped and noised, and the original
evaluation attacked hop~1 and passed its gate. Hop~3 crossed raw and carried the
leak. The zero-support signal is an implementation and system-design defect; the
false pass is an uninstrumented-channel evaluation failure.}
\label{fig:twowire}
\end{figure}

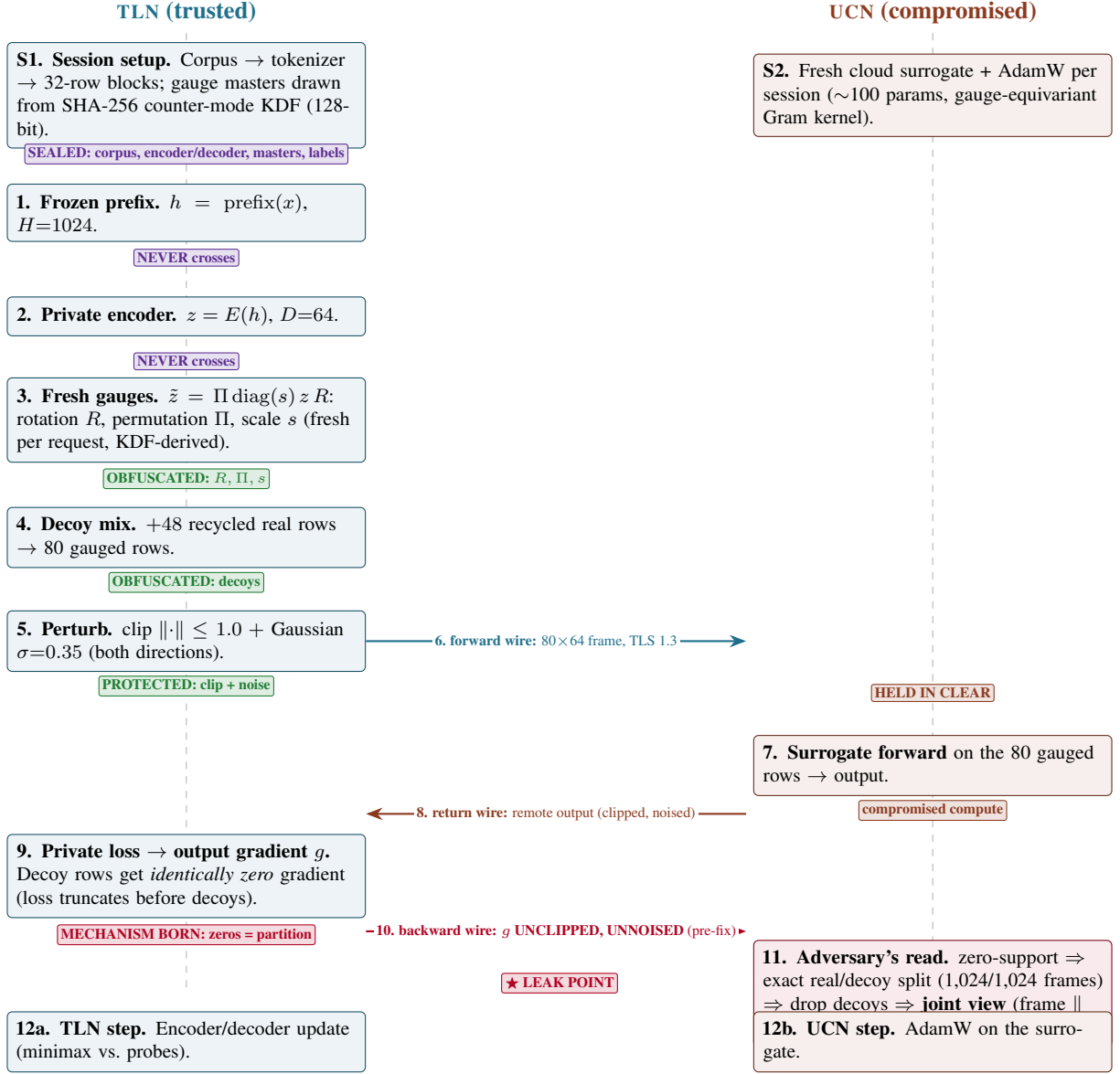
\begin{figure}[H]
\centering
\begin{tikzpicture}[x=1cm,y=1cm]
\node[font=\small\bfseries,text=tlncol] at (2.6,0.3) {\textsc{tln} (trusted)};
\node[font=\small\bfseries,text=ucncol] at (13.4,0.3) {\textsc{ucn} (compromised)};
\draw[dashed,gray!60] (2.6,0) -- (2.6,-13.9);
\draw[dashed,gray!60] (13.4,0) -- (13.4,-13.9);

\stepbox{2.6,-0.9}{setup}{\textbf{S1. Session setup.} Corpus $\to$ tokenizer $\to$ 32-row blocks; gauge masters drawn from SHA-256 counter-mode KDF (128-bit).}{tlncol}
\abadge{sealedcol}{2.6,-1.75}{SEALED: corpus, encoder/decoder, masters, labels}
\stepbox{13.4,-0.9}{ksetup}{\textbf{S2.} Fresh cloud surrogate + AdamW per session ($\sim$100 params, gauge-equivariant Gram kernel).}{ucncol}

\stepbox{2.6,-2.6}{s1}{\textbf{1. Frozen prefix.} $h=\mathrm{prefix}(x)$, $H{=}1024$.}{tlncol}
\abadge{sealedcol}{2.6,-3.25}{NEVER crosses}
\stepbox{2.6,-4.1}{s2}{\textbf{2. Private encoder.} $z=E(h)$, $D{=}64$.}{tlncol}
\abadge{sealedcol}{2.6,-4.75}{NEVER crosses}
\stepbox{2.6,-5.6}{s3}{\textbf{3. Fresh gauges.} $\tilde z=\Pi\,\mathrm{diag}(s)\,z\,R$: rotation $R$, permutation $\Pi$, scale $s$ (fresh per request, KDF-derived).}{tlncol}
\abadge{guardcol}{2.6,-6.45}{OBFUSCATED: $R,\Pi,s$}
\stepbox{2.6,-7.3}{s4}{\textbf{4. Decoy mix.} $+48$ recycled real rows $\to$ 80 gauged rows.}{tlncol}
\abadge{guardcol}{2.6,-7.95}{OBFUSCATED: decoys}
\stepbox{2.6,-8.8}{s5}{\textbf{5. Perturb.} clip $\lVert\cdot\rVert\le 1.0$ $+$ Gaussian $\sigma{=}0.35$ (both directions).}{tlncol}
\abadge{guardcol}{2.6,-9.45}{PROTECTED: clip + noise}

\wirelbl{5.2,-8.8}{10.7,-8.8}{\textbf{6. forward wire:} 80$\times$64 frame, TLS 1.3}{tlncol}
\abadge{ucncol}{13.4,-9.55}{HELD IN CLEAR}

\stepbox{13.4,-10.6}{s7}{\textbf{7. Surrogate forward} on the 80 gauged rows $\to$ output.}{ucncol}
\abadge{ucncol}{13.4,-11.25}{compromised compute}
\wirelbl{10.7,-11.3}{5.2,-11.3}{\textbf{8. return wire:} remote output (clipped, noised)}{ucncol}

\stepbox{2.6,-12.2}{s9}{\textbf{9. Private loss $\to$ output gradient $g$.} Decoy rows get \emph{identically zero} gradient (loss truncates before decoys).}{tlncol}
\abadge{leakcol}{2.6,-13.05}{MECHANISM BORN: zeros = partition}
\wirelbl{5.2,-13.0}{10.7,-13.0}{\textbf{10. backward wire:} $g$ \textbf{UNCLIPPED, UNNOISED} (pre-fix)}{leakcol}
\abadge{leakcol}{8.0,-13.75}{$\bigstar$ LEAK POINT}

\stepbox{13.4,-13.9}{s11}{\textbf{11. Adversary's read.} zero-support $\Rightarrow$ exact real/decoy split (1{,}024/1{,}024 frames) $\Rightarrow$ drop decoys $\Rightarrow$ \textbf{joint view} (frame $\Vert$ gradient).}{leakcol}

\stepbox{2.6,-14.6}{s12a}{\textbf{12a. TLN step.} Encoder/decoder update (minimax vs.\ probes).}{tlncol}
\stepbox{13.4,-14.6}{s12b}{\textbf{12b. UCN step.} AdamW on the surrogate.}{ucncol}
\end{tikzpicture}
\caption{Anatomy of one training cycle, per frame. \textcolor{sealedcol}{SEALED}
never crosses the boundary; \textcolor{guardcol}{OBFUSCATED/PROTECTED} are the
defence's active layers; \textcolor{leakcol}{LEAK} marks the unprotected backward
wire and the partition mechanism it discloses. The forward probe passes at step
6 while steps 10 and 11 leak: the paper's finding in one diagram.}
\label{fig:anatomy}
\end{figure}

The threat model is a fully compromised remote node: \ucn{} can read every
message, state update, and cross-step observation on its own side. Prior split-learning
work demonstrates passive reconstruction and label leakage as well as active
backward-signal manipulation~\cite{erdogan2022unsplit,li2022labelleakage,pasquini2021tiger}.
Accordingly, the declared adversary view has multiple channels, and a privacy
claim is only as strong as the enumeration and testing of those channels.
Figure~\ref{fig:twowire} depicts the protocol's three hops,
Figure~\ref{fig:anatomy} expands one frame into its per-step anatomy, and
Table~\ref{tab:trust} records the trust split at the evaluated operating point.

\begin{table}[H]
\centering\small

\begin{tabular}{@{}p{0.46\linewidth}p{0.46\linewidth}@{}}
\toprule
\textbf{\tln{} (trusted) keeps} & \textbf{\ucn{} (compromised) sees/holds} \\
\midrule
Private corpus; token IDs; plaintext I/O &
Gauged $D{=}64$ frame rows only (80 rows/frame: 32 real $+$ 48 decoy) \\
Frozen LM prefix, embedding, LM tail &
A $\sim$100-parameter gauge-equivariant surrogate it trains itself \\
Private encoder $H{\to}D{=}64$ and decoder $D{\to}H$ weights &
\emph{Unclipped, unnoised} $D$-width output gradients (the omitted channel) \\
Per-request gauge masters (rotation, permutation, scale; SHA-256 counter-mode, 128-bit) &
Protocol metadata: frame sizes, timing, session shape digests \\
Honest evaluation labels; session keys; optimiser state for trusted parts &
TLS~1.3 endpoints and ciphertext (pinned CA) \\
\midrule
\multicolumn{2}{@{}l@{}}{\emph{Never} crosses: $H$-width activations, canonical coordinates, token order, token scale, plaintext tokens.} \\
\bottomrule
\end{tabular}
\caption{Trust split at the evaluated operating point. The output gradient
crosses the backward wire unclipped and unnoised.}
\label{tab:trust}
\end{table}

\paragraph{Terminology.} Four project terms appear throughout without prior-art
homes, so we define them once here (no citations; they are constructs of this
system, not borrowed results). A \emph{seed} is the random seed initialising one
training run and, through the KDF, every per-request gauge draw --- the unit of
independent replication. A \emph{cell} is one complete experiment --- a defence
configuration, a seed, and a training budget, run end-to-end and then attacked
and scored. A \emph{battery} is a predeclared set of attacks scored as one
sweep: the frozen nine-arm probe family (three model classes $\times$ three
restarts) or the compromise-fraction sweep. A \emph{surrogate} is the small
gauge-equivariant module \ucn{} trains in place of the real middle layers
($\sim$100--161 parameters) --- a stand-in that computes on gauged frames
without ever learning the gauges. Likewise: the \emph{gate} is the predeclared
decision threshold, the \emph{floor} is the reading of a matched no-attack
control, an \emph{arm} is one scored attacker instance, \emph{chaff} is the
recycled real decoy rows, and a \emph{gauge} is one fresh per-request
randomisation (rotation, permutation, or scale).

\section{Evaluation protocol: declared channels, calibrated metrics, and the
gate statistic}
\label{sec:protocol}

The gap this paper addresses is not discovery of gradient leakage or a
bidirectional attack surface: both are established in gradient-inversion and
split-learning work~\cite{zhu2019deep,deng2021tag,bisr}. The gap is a concrete
evaluation returning a pass without testing whether its metrics can detect a
known leak on every declared channel. We instantiate three established audit
disciplines for that setting.

\paragraph{Declare every channel.}
The evaluation must enumerate all channels the adversary observes (forward,
backward, membership, timing) before it measures any of them. A channel absent
from the declared view is exempt from the gate by construction; the leak found
here lived exactly in that exemption. Table~\ref{tab:channels} records the enumeration and its current status. Its
artefact sources and verification procedures are indexed in
Appendix~\ref{app:artifacts}.

\begin{table}[H]
\centering\scriptsize
\begin{tabular}{@{}p{0.20\linewidth}p{0.30\linewidth}p{0.26\linewidth}p{0.16\linewidth}@{}}
\toprule
family & applicable metrics & positive control & status \\
\midrule
\texttt{forward\_only} & \texttt{token\_top1}, \texttt{rare\_token\_top1}, \texttt{token\_cross\_entropy} & injected leak on the forward wire; codeword injection & measured \\
\texttt{gradient\_only} & \texttt{token\_top1}, \texttt{rare\_token\_top1}, \texttt{token\_cross\_entropy} & injected\_leak on the backward wire & measured \\
\texttt{joint\_forward\_gradient} & \texttt{token\_top1} & scaled joint concatenation (the exploratory frequent-token gradient result) & measured \\
\texttt{accumulated\_history} & \texttt{token\_top1} & none & unmeasured (no positive control) \\
\texttt{stateful\_remote\_state} & \texttt{token\_top1} & none & unmeasured (no positive control) \\
\texttt{timing\_metadata} & --- & none & unmeasured (no applicable metric) \\
\texttt{active\_perturbation} & \texttt{token\_cross\_entropy} & amplitude sweep & constructible, unexecuted \\
\texttt{membership\_property} & --- & none & unmeasured (no applicable metric) \\
\texttt{response\_side} & --- & none & unmeasured (no applicable metric) \\
\bottomrule
\end{tabular}

\caption{The declared channel families and their status. ``Unmeasured'' means the family has no applicable metric or no positive
control, so it cannot support a primary claim.}
\label{tab:channels}
\end{table}

\paragraph{Calibrate every metric against a known leak.}
A metric certifies the absence of a leak only if it detects a leak injected on
purpose. Section~\ref{sec:calibration} measures each metric's detection threshold with a controlled
dose--response sweep. A metric that has not been calibrated cannot stand between
a privacy claim and a passing verdict. Planted canaries and attack-based privacy
audits provide the methodological precedent~\cite{carlini2019secret,jagielski2020auditing,nasr2023tight,steinke2023onerun,pillutla2023randomization};
our adaptation makes the dose and decision channel- and metric-specific.

\paragraph{Define the gate statistic and paired effects.}
Every attack result in this paper is a top-1 token accuracy: the share of
evaluation tokens the attacker's probe names correctly. We report it not as a
raw accuracy but as the gap, in percentage points (pp), between the probe and a
constant baseline that always guesses the single most frequent token in the
evaluation set. That baseline sits near 5 to 6\% for this model and corpus, so
an effect of $+1$~\pp{} means the attacker recovers roughly one extra token per
hundred, about a sixth more than guessing alone would give. The gate is set at
$+1.0$~\pp{}, and the effects measured here fall between about $+0.5$ and
$+2.3$~\pp{}.

Let $\mathcal J$ be the nine predeclared probe arms (three probe variants,
each scored at three restarts; Table~\ref{tab:setup}),
$U^{\mathrm{Bonf}}_{0.95}(\widehat p_j)$ the Bonferroni-adjusted Wilson upper-95
accuracy for arm $j$, and $\widehat p_{\mathrm{const}}$ the point accuracy of
the constant baseline, which always predicts the most frequent evaluation
token. The historical gate statistic is
\begin{equation}
G = \max_{j\in\mathcal J} U^{\mathrm{Bonf}}_{0.95}(\widehat p_j)
    - \widehat p_{\mathrm{const}} .
\end{equation}
Thus the constant baseline is subtracted \emph{after} the Wilson bound is
formed; $G>+1.0$~\pp{} fails the gate. For frame $f$, define the paired
difference
\begin{equation}
\Delta_f(q)=\operatorname{acc}_f(q)-
             \operatorname{acc}_f(\text{constant baseline}).
\end{equation}
For a real arm $r$ and its shuffled-label negative control $s$, respectively,
\begin{equation}
A = \frac{1}{F}\sum_{f=1}^{F}\Delta_f(r),
\qquad
N = \frac{1}{F}\sum_{f=1}^{F}\Delta_f(s).
\end{equation}
$A$ is the real-arm paired effect; $N$ is the same paired effect for the
shuffled-label negative control. $N$ is a false-positive check and is
\emph{not} subtracted inside $G$. Within-run uncertainty bootstraps frames;
reported across-seed contrasts, including $A-N$, use a hierarchical bootstrap
over seeds (outer) and frames (inner).

Three verdict terms recur throughout, and we fix them here. A result is
\emph{detected} when its lower bootstrap bound is positive or when exact
support evidence establishes it; an arm is \emph{at floor} when that bound
includes zero, so the arm is indistinguishable from its constant baseline; and
an arm \emph{breaks the gate} when, and only when, $G > +1.0$~\pp{} under the
Bonferroni--Wilson rule of Equation~(1). ``Detected'' and ``breaks the gate''
are not synonyms: a paired effect can be detected well below the gate, and the
gate can break on an upper bound whose point effect is smaller.
We also call one (configuration, seed) run a \emph{cell}, and one capture run's
serialised frames plus its metadata manifest a \emph{bundle}.

$z$-style row-independent statistics remain
only for comparability with previously reported results of this system. Calibrated reference
distributions and explicit operating points are established in
membership-inference evaluation~\cite{carlini2022lira,song2021systematic}; the
shuffled-label negative control is their channel-specific counterpart here.

\section{Instrument calibration}
\label{sec:calibration}

Calibration is per metric: thresholds are set only after each metric's
detection curve is measured. The sweep injects a known token leak at controlled
dose (coverage $\times$ amplitude) and reads where each metric detects it.

\subsection{The four metrics disagree}

\begin{figure}[H]
\centering
\includegraphics[width=\linewidth]{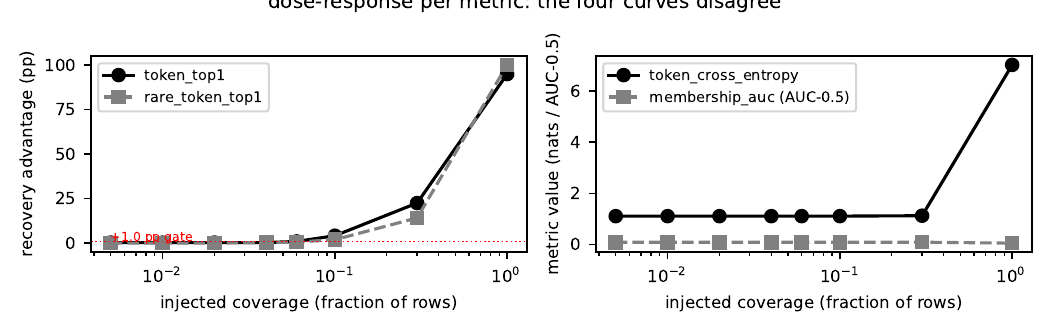}
\caption{Dose--response per metric on the coordinate-mode sweep (amplitude
1.0). The four emitting metrics disagree: \texttt{rare\_token\_top1} is most
sensitive; \texttt{token\_cross\_entropy} barely moves until the leak dominates;
\texttt{membership\_auc} does not detect the low-dose token signal because the
injection is not a membership signal.
A single calibration run across 19 doses.}
\label{fig:dose}
\end{figure}

Figure~\ref{fig:dose} and Table~\ref{tab:dose} show the dose--response curves.
\texttt{token\_top1} has a sharp onset between coverage 0.04 and 0.06, steepening
through 0.10. \texttt{rare\_token\_top1} is the most sensitive, first responding
at coverage 0.04 and reaching full recovery at coverage 1.0.
\texttt{token\_cross\_entropy} is dose-insensitive until the injected leak
dominates. \texttt{membership\_auc} is insensitive over the tested low- and
moderate-dose region (AUC$-0.5$ spans
$0.058$--$0.159$ across the full sweep), because the injected leak is a
token-identity signal, not a membership signal; the metric was subsequently
falsified as a channel and retained as a probe-generalisation diagnostic only.

The curves disagree, consistent with broader evidence that reconstruction
metrics need not agree on privacy risk~\cite{sun2023metrics}, so no single
threshold fits all metrics. The thresholds (Table~\ref{tab:thresholds}) are set per metric, and they are budget-
and frame-invariant along the remaining sweep axes. Calibration changed how the
decision is interpreted, not the gate formula: three cells later shown to be
degenerate (arm identical to the constant baseline row-for-row) had pinned the
statistical floor, and the sweep measures that false-negative region while
preserving the historical $+1.0$~\pp{} Bonferroni--Wilson rule. The frequent-token
effect is detected on every seed ($A$ from $+0.69$ to $+1.19$~\pp{}, $|N|\le
0.08$~\pp{}); explicit gate breaks are reported only where the
Bonferroni--Wilson column exceeds $+1.0$~\pp{} (Section~\ref{sec:results}).

\begin{table}[H]
\centering\small
\begin{tabular}{@{}lrrrrrr@{}}
\toprule
metric & cov 0.02 & 0.04 & 0.06 & 0.10 & 0.30 & 1.00 \\
\midrule
\texttt{token\_top1} (pp)        & 0.35 & 0.35 & 0.98 & 4.19 & 22.51 & 94.88 \\
\texttt{rare\_token\_top1} (pp)  & 0.08 & 0.32 & 0.67 & 1.97 & 14.18 & 100.0 \\
\texttt{token\_cross\_entropy}   & 1.114 & 1.114 & 1.114 & 1.115 & 1.130 & 7.025 \\
\texttt{membership\_auc} (AUC$-0.5$) & 0.093 & 0.093 & 0.095 & 0.095 & 0.096 & --- \\
\bottomrule
\end{tabular}

\caption{Coordinate-mode coverage sweep at amplitude 1.0. The calibration reference is the zero-dose
source cell (coverage 0, same bundle family) and the constant baseline (the most
common evaluation token); the
calibration sweep contains no shuffled-label arm, because those negative
controls are introduced only at scoring time. The top-1 rows show point effects, while the
gate statistic uses
their Bonferroni--Wilson upper-95 value; thus the gate first breaks at coverage
0.06 although the displayed point effect is $0.98$~\pp{}. Cross-entropy is a
proper scoring rule; the values shown are the effect over the constant
baseline, and at scoring time CE is read arm minus its shuffled-label negative
control, never raw
(Table~\ref{tab:thresholds}).
\texttt{membership\_auc} values are flat. ``---'' marks the amplitude-axis
section, not part of this row.}
\label{tab:dose}
\end{table}

\begin{table}[H]
\centering\scriptsize
\begin{tabularx}{\textwidth}{@{}>{\raggedright\arraybackslash}p{3.2cm}>{\raggedright\arraybackslash}X>{\raggedright\arraybackslash}X@{}}
\toprule
metric & threshold set in advance & basis \\
\midrule
\texttt{token\_top1} & gate: Bonferroni upper-95 excess $>+1.0$ pp & first displayed gate break at cov.\ 0.06 \\
\texttt{rare\_token\_top1} & paired effect $+1.0$ pp; report raw recovery & earliest subthreshold movement at cov.\ 0.04 \\
\texttt{token\_cross\_entropy} & none set; read arm minus shuffled-label negative control (never raw) & a confidently-wrong negative control inverts raw CE; no numerical gate \\
\texttt{membership\_auc} & none (diagnostic only) & falsified as a channel; constant baseline vacuous \\
\bottomrule
\end{tabularx}

\caption{Per-metric thresholds, set in advance. A
metric without a measured detection curve may not appear in a primary claim.}
\label{tab:thresholds}
\end{table}

\subsection{The representation-matched positive control}
\label{sec:positive_control}

A floor reading is interpretable only if the same probe succeeds on a
representation-matched undefended control. The isolation-audit bundles do not
provide that control because their naked boundary is a different
representation and the probe remained near floor there. For the split-14
claims, the positive control therefore uses the split-14 naked capture and a
probe built to detect it: a four-layer, eight-head Transformer encoder trained
50 epochs to map each released latent row back to its token
. Appendix~\ref{app:artifacts}
indexes the associated artefacts.

\begin{table}[H]
\centering\small
\begin{tabular}{@{}lcccl@{}}
\toprule
boundary & Wilson upper-95 acc.\ & constant baseline & excess & note \\
\midrule
naked ($D{=}1024$, no defence) & \textbf{28.54\%} & 4.35\% & \textbf{$+24.19$ pp} & breaks \\
defended ($D{=}64$, noise, decoys) & 5.56\% & 5.27\% & $+0.29$ pp & at floor \\
defended ($D{=}1024$, dimension-matched) & 5.56\% & 5.29\% & $+0.27$ pp & at floor \\
\bottomrule
\end{tabular}

\caption{One probe architecture, one protocol, three boundaries. The probe that
inverts the naked boundary at over six times its baseline reads nothing through the
defence, at the operating width \emph{and} at matched width, so the
defended floor readings are not attacker blindness. Best accuracy is selected
on the evaluation set over 50 epochs (a selection reading); the accuracy column
reports its Wilson upper-95 bound, so each row's excess is the displayed accuracy
minus the baseline. The defended probes' accuracy declines after its early epochs,
consistent with overfitting noise rather than extracted signal. The
committed three-restart rerun of the $D{=}64$ defended probe reads at the same
floor ($+0.37$~\pp{}); the other three committed deep-probe runs are the rows
above.}
\label{tab:positive_control}
\end{table}

The naked break selects the best epoch on the evaluation set,
so $+24.19$~\pp{} is a sensitivity demonstration, not a primary leak estimate; and
the defended frame carries 80 rows against the naked frame's 32, matching the
study's evaluation convention. What the contrast establishes is narrow and
load-bearing: an attacker architecture proven sensitive to the representation
reads nothing through the defence, while an attacker family that could not read
even the naked boundary could never have shown it. Boundary-specificity is measured: re-run on the packaged isolation-audit bundle, the same probe reads
the naked boundary at floor as well ($+0.66$~\pp{}, its shuffled-label
negative control at $+0.50$). This is exactly why a
positive control must be representation-matched to the cell under test, and why
the split-14 capture above is the control for the split-14 claims.

\section{Results: structural gradient leakage and bounded content inference}
\label{sec:results}

Table~\ref{tab:summary} summarises the main configuration and the result
categories developed in this section; availability and verification details
are indexed in Appendix~\ref{app:artifacts}.

\begin{table}[H]
\centering\small
\begin{tabularx}{\textwidth}{@{}>{\raggedright\arraybackslash}p{2.3cm}>{\raggedright\arraybackslash}X>{\raggedright\arraybackslash}X@{}}
\toprule
result category & result & interpretation \\
\midrule
forward wire & gate passed & the original instrument's only channel \\
backward wire & leaks structurally & 4,096/4,096 frames; row agreement 1.000, on
  each of nine seeds \\
frequent-token content effect & pooled over the nine verified seeds:
  $+0.92$~\pp{}, 95\% interval $[0.74, 1.09]$ & negative control at floor on every
  seed; per-seed values in Table~\ref{tab:sixseed} \\
nine audit cells & effect detected at 12 and 11 delegated layers, at floor at
8 and 6; detected at $D{=}64$ and $D{=}96$, at floor at $D{=}128$; 2.5$\times$
more exposure does not amplify & one run each (seed 42); partition exact at
every depth \\
dose--response calibration & \texttt{token\_top1} onset $\sim$0.04--0.06;
  \texttt{rare\_token\_top1} at 0.04; \texttt{CE} flat until dominant;
  \texttt{membership\_auc} flat and dose-insensitive & per-metric thresholds
  set in advance \\
internal scorer reimplementation & all nine cells $\le 10^{-6}$ pp &
  internal check (Appendix~\ref{app:artifacts}) \\
\midrule
mitigation runs & joint view breaks the gate on $6/6$ defended cells across
  two datasets; gradient clipping and Gaussian noise suppress the scored probes
  for a held-out cost of $\approx 0.01$ nats & under the fixed protocol; four-layer topology \\
\bottomrule
\end{tabularx}

\caption{Result categories and where each is established. The final row refers
to the four-layer mitigation topology; every row above it refers to the primary
configuration.}
\label{tab:summary}
\end{table}

\subsection{Mechanism: the gradient says which rows are decoys}
\label{sec:mechanism}

A decoy row's gradient is identically zero, because the trusted-side loss
truncates before the decoys. With the gradient unprotected, the zero-support
pattern of each returned gradient frame exactly repeats the split between
real rows and decoys. On every one of 4,096 frames, on every seed, the match is exact:
\textbf{4,096/4,096 frames; row-level agreement 1.000}. The corresponding
per-seed artefacts are indexed in Appendix~\ref{app:artifacts}.

The result has three claim levels.
\paragraph{(i) Structural metadata leakage.}
Gradient support discloses exactly which rows are real: the cloud can
separate every loss-bearing row from every decoy row in every evaluated frame.
This is the disclosure the backward wire is uniquely responsible for.
\paragraph{(ii) Content inference.}
Conditioned on that structural signal, the pre-set frequent-token probe has a
modest paired effect. The gradient's contribution here is small: comparing the
gradient and forward arms of Table~\ref{tab:sixseed} seed by seed, the
gradient's margin over the oracle-partitioned forward frame is at most
$+0.18$~\pp{} on five of the six exploratory seeds ($+0.0011$ to $+0.0395$
on four of them), and reaches $+0.64$~\pp{} only on seed 42. The content an attacker reads therefore sits largely in the
forward frame, and what the backward wire supplies is the partition that makes
the frame readable, which is claim~(i), not additional content. This probe
result is content inference; it is not evidence of transcript or sequence
reconstruction.
\paragraph{(iii) Not established.}
The experiments do not establish rare-token recovery, sequence reconstruction,
or held-out-text reconstruction. Those claims require different emitters and
controls and remain outside the measured result.

The cycle in which this happens is Figure~\ref{fig:anatomy}, steps 9 to 11.
Table~\ref{tab:sixseed} reports the exploratory and replication runs.

\begin{table}[H]
\centering\small
\begin{tabular}{@{}lccccc@{}}
\toprule
seed & gradient arm (pp) & forward arm (pp) & negative control (pp) & frames exact & verdict \\
\midrule
42 & $+1.1923$ & $+0.5516$ & $-0.0069$ & \textbf{4,096/4,096} & detected \\
43 & $+0.6893$ & $+0.6821$ & $-0.0231$ & \textbf{4,096/4,096} & detected \\
44 & $+0.6929$ & $+0.6918$ & $+0.0479$ & \textbf{4,096/4,096} & detected \\
45 & $+0.8435$ & $+0.8040$ & $-0.0707$ & \textbf{4,096/4,096} & detected \\
46 & $+0.9361$ & $+0.7578$ & $-0.0285$ & \textbf{4,096/4,096} & detected \\
47 & $+0.7705$ & $+0.7618$ & $-0.0308$ & \textbf{4,096/4,096} & detected \\
\midrule
mean & $+0.8541$ & $+0.7082$ & $|\cdot| \le 0.08$ & \textbf{4,096/4,096} & --- \\
\midrule
48 (fixed protocol) & $+0.6456$ & $+0.5023$ & $-0.0571$ & \textbf{4,096/4,096} & detected \\
49 (fixed protocol) & $+1.5008$ & $+0.8721$ & $+0.0232$ & \textbf{4,096/4,096} & detected \\
50 (fixed protocol) & $+0.9929$ & $+0.6326$ & $+0.0650$ & \textbf{4,096/4,096} & detected \\
\bottomrule
\end{tabular}

\caption{Exploratory seeds 42--47 and the replication runs (seeds
48--50). Statistic: the paired effect of each arm over its constant baseline
(the most frequent evaluation token), clustered by frame. The negative-control
column reports the label-shuffled twin of the gradient arm, scored through the
same statistic. Inferential unit: one training seed; each row one independent
run, its interval a cluster bootstrap over that run's frames. All nine
committed seeds are shown. \emph{The forward arm is built on the oracle
real/decoy split} and therefore does not by itself describe an achievable
attack. It is reported here because in this system the oracle is redundant:
the gradient reproduces the partition exactly (Section~\ref{sec:mechanism}), so
the forward arm is what an attacker reads \emph{after} the backward wire has
supplied the partition. No label-shuffled control was run for the forward arm,
so that column is uncontrolled; the negative-control column bounds the gradient
arm only. Seed 42 was served by a cloud container built before the code was
packaged for release; seeds 43--47 ran the released package on both nodes, and
seeds 48--50 are captures made under the fixed protocol.}
\label{tab:sixseed}
\end{table}

\subsection{What this does, and does not, establish}

The exact partition and the modest frequent-token paired effect are detected
under a shuffled-label negative-control protocol the audited evaluations do not
run. It establishes that a gate that never
declares the backward channel cannot see the partition signal. The originally
published gate would also have passed this cell under its own threshold. The
structural disclosure was invisible to it twice over: the channel was never
scored, and the floor the threshold was calibrated against was degenerate
(Section~\ref{sec:calibration}).

\subsection{Does the leak survive a configuration worth deploying?}
\label{sec:confirmatory}

The main configuration fails its own utility gate: it protects the data but
costs too much model quality to be worth running. That leaves an objection.
If the leak only appears in a configuration nobody would deploy, it may not
matter. This section answers that objection with a second set of runs on a
configuration that does pass both gates, so the leak cannot be dismissed as an
artefact of an unusable setting.

These runs use a shallower split, delegating four transformer layers instead of
eleven. They were run on three fresh seeds (51 to 53) from the released code,
and scored against the thresholds set in advance, so nothing about them was
tuned after the fact.

Table~\ref{tab:confirmatory} reports every run in the mitigation set.

\begin{table}[H]
\centering\scriptsize
\setlength{\tabcolsep}{3pt}
\begin{tabularx}{\textwidth}{@{}>{\raggedright\arraybackslash}p{3.2cm}*{5}{>{\centering\arraybackslash}X}@{}}
\toprule
cell (seed) & utility $\Delta$ & forward gate & joint arm (gate) & joint arm (paired effect) & support leak \\
\midrule
defended, gradient open (s51) & $+0.054$ & $+0.522$ & \textbf{$+1.954$} & $+1.609$ & \textbf{1,024/1,024} \\
defended, gradient open (s52) & $+0.063$ & $+0.574$ & \textbf{$+2.180$} & $+2.333$ & \textbf{1,024/1,024} \\
defended, gradient open (s53) & $+0.071$ & $+0.412$ & \textbf{$+1.765$} & $+1.567$ & \textbf{1,024/1,024} \\
\addlinespace
defended, gradient clipped and noised (s51) & $+0.063$ & $+0.406$ & $+0.480$ & $-0.204$ & \textbf{0/1,024} \\
defended, gradient clipped and noised (s52) & $+0.078$ & $+0.421$ & $+0.476$ & $-0.017$ & \textbf{0/1,024} \\
defended, gradient clipped and noised (s53) & $+0.083$ & $+0.412$ & $+0.481$ & $-0.013$ & \textbf{0/1,024} \\
\addlinespace
public corpus, gradient open (s51) & $+0.092$ & $+0.385$ & \textbf{$+1.632$} & $+0.986$ & \textbf{1,024/1,024} \\
public corpus, gradient open (s52) & $+0.080$ & $+0.767$ & \textbf{$+1.607$} & $+0.942$ & \textbf{1,024/1,024} \\
public corpus, gradient open (s53) & $+0.083$ & $+0.483$ & \textbf{$+1.435$} & $+0.750$ & \textbf{1,024/1,024} \\
\addlinespace
naked control (s51--s53) & $+0.00/+0.12/+0.43^{\dagger}$ & $+9.5$ to $+10.7$ & $+33.9$ to $+34.0$ & $\approx +51$ & --- \\
\bottomrule
\end{tabularx}

\caption{The mitigation runs. The gate columns report
the gate statistic: Bonferroni-upper-95 excess over the constant baseline. The
paired column is the cluster-bootstrapped paired effect of the best arm against
its constant baseline. ``Joint arm'' is the forward frame and the output
gradient of the same training step, concatenated: the view the compromised node
actually holds, and the one the original gate never scored. This set
comprises fourteen runs; the twelve scored cells are shown. The two omitted are
a configuration smoke test that produced no scored arms and the packaged
seed-42 rerun of the main configuration, reported separately in this
section. $^{\dagger}$The naked control on seed 53 fails the utility gate
($\Delta$ loss $0.432$ against the $0.35$ bound); it is retained because it is
a deliberately undefended sensitivity control, not a candidate configuration.
\emph{In every one of these runs the per-row scale-and-sign gauge was
disabled}, as it is for the rotation-invariant cloud architectures
(Appendix~\ref{app:stack}); the joint-view results below are conditional on
that setting.}
\label{tab:confirmatory}
\end{table}

(a)~\textbf{The joint-view gate result is not confined to a utility-failing
configuration.} On all six defended, gradient-open cells (three seeds on each
of two datasets) the run passes the forward privacy gate \emph{and} the
utility gate, and the joint view still breaks the $+1.0$~\pp{} gate set in advance,
while the gradient on its own stays at the level of guessing.

(b)~\textbf{The mitigation removes the zero-support partition signal.} With
per-row gradient clipping and Gaussian noise, every scored arm sits at floor
and the structural partition disclosure is gone: 0/1,024 frames, agreement
$0.400$ at the 32/80 base rate. Utility is intact, at a held-out cost of
$\approx 0.01$~nats.

(c)~\textbf{The naked control restores the signal.} The same topology with the
defence off reads $+34$~\pp{} on the joint view, so the defended cells'
near-floor readings reflect the defence rather than a blind instrument. On the
forward gate the defended cells sit at floor while the naked cells exceed it by
roughly ten \pp{}, separating the two configurations by more than an order of
magnitude.

\textbf{Cross-Gram cancellation does not explain the joint-view result.} The forward
activation matrix $X$ and its returned gradient $G$ share the per-request
rotation, so their cross-Gram product $XG^{T}$ cancels it. The pre-set probe family did not compute this cross-tensor feature. We implemented it (cosine and
scale-keeping forms, with
rotation-invariance self-tests) and scored it on every captured cell: it reads
at floor everywhere, defended (gate statistic $+0.31$ to $+0.86$, paired effect
$-0.29$ to $+0.46$) and naked alike (paired effect $-0.78$ to $-0.22$). Under
the two implemented emitters, the joint-view effect therefore runs through the
concatenated gradient block's own content, not through the rotation-cancelling
cross-term; other cross-tensor attacks remain open.

\textbf{The effect exists on a second corpus.} The main results are on the
private WikiText-2 slice. The alternative public WikiText-2 corpus carries the
same topology on seeds 51--53 (Table~\ref{tab:confirmatory}), and on two of the
three the gradient alone breaks the gate as well ($+1.175$ and $+1.219$). One
additional corpus is a limited robustness check: the effect is observed on both
tested datasets, at different magnitudes, which does not establish that the
effect is dataset-independent.

\textbf{Packaged seed-42 rerun.} The seed-42 result in
Table~\ref{tab:sixseed} ($+1.1923$, originally served by a pre-existing
container) was re-run from packaged code in a clean container in the primary
configuration: $+1.147$ (upper-95), paired effect $+1.017$, support exact on
4,096/4,096, and the negative control at floor. The result is not confined to
that container.
Pooling the nine seeds that were run from packaged code (the six exploratory
and the three replication runs), a seed-level random-effects estimate
puts the frequent-token gradient-arm paired effect at $+0.92$~\pp{}, 95\%
interval $[0.74, 1.09]$ ($\tau = 0.23$, $I^2 = 77\%$). The interval reaches the
$+1.0$~\pp{} gate, so the pooled content effect is not separable from the gate
threshold at this sample size. Three further units from an earlier, unpackaged
tree are retained in the committed estimate for continuity with earlier
reports; including them gives $+0.88$~\pp{}, $[0.75, 1.02]$, but they are not
re-derivable from the release and we do not rely on them.

These runs are 2{,}000-step diagnostics on the
0.6B model, and the clean 40k-step rerun confirms the main
configuration still fails its utility gate ($+0.896$, vs $0.9185$ on the
original cell). The claim that a
configuration passes both gates attaches to the four-layer topology, not to the
eleven-layer main configuration.

\section{When does the structural signal convert to a token advantage?
Depth, width, and budget}
\label{sec:shape}

\begin{figure}[H]
\centering
\includegraphics[width=\linewidth]{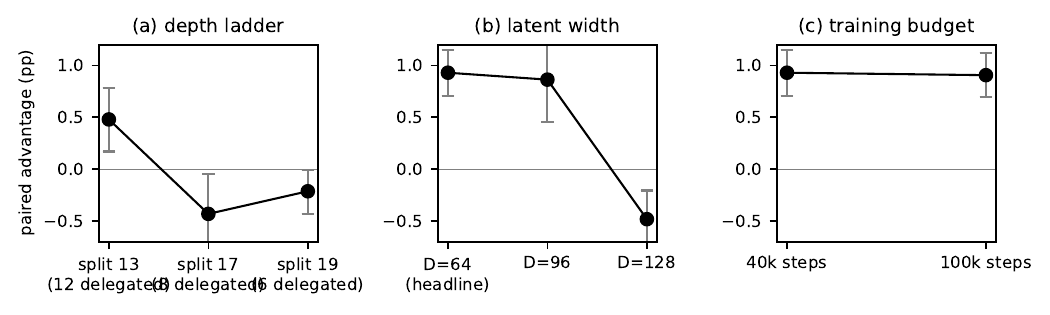}
\caption{Attack-specific shape across audit cells. (a) depth ladder: the
implemented emitter detected the effect at 12--11 delegated layers and not
at 8--6. (b) the width sweep: the effect is detected at $D{=}64$ and $D{=}96$, and not
at $D{=}128$.
(c) The implemented probe's paired effect did not increase between the two
tested exposure budgets. Paired effect over the constant baseline with 95\% CI
on the gradient arm; one run per cell
(seed 42).}
\label{fig:shape}
\end{figure}

Nine audit cells vary budget, depth, and width
(Figure~\ref{fig:shape}; Table~\ref{tab:shape}); each cell is a single run on
seed 42, so every shape threshold below is a single-seed reading. Within this
run the readings were insensitive to budget between 40k and 100k steps. Depth and
width thresholds are read on the paired effect, and every cell's
shuffled-label negative control reads at floor.

\begin{table}[H]
\centering\small
\begin{tabular}{@{}llrrcl@{}}
\toprule
axis & steps & gradient arm (pp) & negative control (pp) & verdict & frames \\
\midrule
depth (12 del.) & 2k & $+0.4796$ & $-0.0702$ & \textbf{detected} & 512/512 \\
depth (11 del., $D{=}64$) & 40k & $+0.9302$ & $-0.0112$ & \textbf{detected} & 4,096/4,096 \\
budget & 100k & $+0.9066$ & $-0.0926$ & \textbf{detected} & 4,096/4,096 \\
width ($D{=}96$) & 10k & $+0.8635$ & $-0.0061$ & \textbf{detected} & 4,096/4,096 \\
depth (8 del.) & 2k & $-0.4314$ & $+0.0399$ & at floor & 512/512 \\
depth (6 del.) & 2k & $-0.2131$ & $+0.0270$ & at floor & 512/512 \\
width ($D{=}128$) & 10k & $-0.4823$ & $-0.0153$ & at floor & 4,096/4,096 \\
width ($D{=}96$) & 10k & $-0.0861$ & $+0.0824$ & at floor & 512/512 \\
width ($D{=}128$) & 10k & $-0.1190$ & $+0.0702$ & at floor & 512/512 \\
\multicolumn{6}{@{}l}{\small(the 512-frame $D{=}96/128$ readings are superseded by the 4,096-frame re-runs)} \\
\bottomrule
\end{tabular}

\caption{The nine-cell audit. Paired effect of the gradient arm over its
constant baseline, with each cell's shuffled-label negative control beside it; frame-exact counts from the zero-support partition check. All nine committed cells, one run each (seed 42). The partition mechanism (zero-support match) is exact on
every cell, whether or not the token effect is detected.}
\label{tab:shape}
\end{table}

\paragraph{Structural, not statistical.}
The split is disclosed on every frame because every decoy gradient is
exactly zero; the disclosure is deterministic, not a sampling artefact.

\paragraph{Exposure budget.}
The implemented probe's paired effect did not increase between the two tested
exposure budgets. From 40k to 100k gradient exposure (2.5$\times$), the
per-row effect changes from $+0.9302$ to $+0.9066$ in this run.

\paragraph{Attack-specific depth and width dependence.}
The recovered token advantage is detectable at 12 and 11 delegated layers and
undetectable at 8 and 6: the gate is bracketed between 8 and 12 on the matched
2k-step ladder, and the 11-layer point is the 40k-step main configuration, where
depth is confounded with budget; the ladder's point estimates are not a
monotone dose--response. The paired effect is detected at latent width $D{=}64$ and $D{=}96$, and not
at $D{=}128$. The
partition classification itself is exact at every
depth and width (Table~\ref{tab:shape}): depth and width affect conversion of
the partition into a token advantage for this configuration and attack, not the
underlying zero-support signal. Stronger inversion attacks can recover prompts
from deep LLM states~\cite{dong2025depth}, so this local transition is not
evidence that depth generally provides privacy.
Table~\ref{tab:attacker_ledger} collects every attack family tried against the
system, where each applies in the training cycle, and its outcome.

\begin{table}[H]
\centering\scriptsize
\setlength{\tabcolsep}{3pt}
\begin{tabular}{@{}p{0.28\linewidth}p{0.08\linewidth}p{0.44\linewidth}p{0.13\linewidth}@{}}
\toprule
attack & step (Fig.~\ref{fig:anatomy}) & outcome & does it work? \\
\midrule
pre-set 9-arm battery on forward frames & 6 & defended floor $+0.3..+0.9$~\pp{}; naked $+9.5..+10.7$~\pp{} & no, against the defence \\
deep-inversion transformer probe & 6 & naked split-14 $+24.19$~\pp{}; defended floor at $D{=}64$ \emph{and} $D{=}1024$ & no, against the defence \\
zero-support read (no training; just zeros) & 10 & partition exact on \emph{every} frame, every seed & \textbf{yes} \\
gradient arm alone & 10 & floor on the private corpus ($+0.50$); breaks on 2 of 3 public-corpus seeds ($+1.18/+1.22$) & only on one dataset \\
joint arm (frame $\Vert$ gradient) & 6$+$10 & breaks the $+1.0$~\pp{} gate set in advance on \textbf{6 of 6} defended cells ($+1.44..+2.18$) & \textbf{yes} \\
$XG^{T}$ cross-Gram invariant & 6$+$10 & floor on every capture (cosine and raw forms) & no; the feature carries nothing \\
vector-matching and nearest-signature attacks & 6 & $0.02$--$0.03\%$ and $1.5\%$ against a $\approx 5.9\%$ constant baseline, i.e.\ below the no-attack control & no \\
compromise battery (wire capture, decoy identities, gauge compromise, known plaintext) & 6 & flat at/below control on every fraction & no \\
\midrule
after the mitigation (per-row clip $0.01$ and Gaussian noise) & 10 & all arms at floor ($+0.48$); support $0/1{,}024$; utility cost $+0.01$~nats & \textbf{no longer} \\
\bottomrule
\end{tabular}

\caption{The attacker ledger: every attack family tried against the system,
where it applies in the cycle of Figure~\ref{fig:anatomy}, and its outcome.
Every committed attack family is listed; the supporting artefacts are indexed
in Appendix~\ref{app:artifacts}. The two leak rows are the finding; the closed
row is the fix.}
\label{tab:attacker_ledger}
\end{table}

\section{External audits and related work}
\label{sec:related}

\subsection{A selected audit of split-LLM evaluations}

The audit selected three recent evaluations to span a bidirectional attack, a
bidirectional defence, and recent attack-plus-defence work; it is a purposeful
sample, not a field survey (Table~\ref{tab:external}). None of the three
combined all three controls.

\begin{table}[H]
\centering\small
\begin{tabularx}{\textwidth}{@{}l l X X X@{}}
\toprule
system & role & injected calibration & shuffled-label negative control & predeclared gate \\
\midrule
BiSR~\cite{bisr} & attack & none & none & none \\
DualGuard~\cite{dualguard} & defence & none & none & none \\
Prompts to Responses~\cite{promptstoresponses} & attack + defence & none & no (unmatched baseline) & none \\
\bottomrule
\end{tabularx}

\caption{Selected audit verdicts. The three works
report attack or defence performance against reconstruction baselines.
``Prompts to Responses'' includes an
unmatched random-token baseline in its appendix.}
\label{tab:external}
\end{table}

BiSR~\cite{bisr} demonstrates backward-channel reconstruction against
perturbation defences through comparisons with re-adapted attacks, but does not
define an acceptance gate or shuffled-label negative control.
DualGuard~\cite{dualguard} correctly
evaluates forward, backward, and bidirectional attack paths before reporting a
worst-case aggregate; its defence comparisons nevertheless do not include an
injected detector canary, shuffled-label negative control, or predeclared
privacy gate. Thus they
do not support the specific calibrated no-leak decision required here; this is
a statement about evaluation semantics, not defence efficacy. From Prompts to
Responses~\cite{promptstoresponses} is forward-only by scope and reports an
unmatched random-token baseline in its appendix.

\subsection{Position within the differential-privacy auditing lineage}

Planted-secret exposure tests~\cite{carlini2019secret} and attack-based DP
audits~\cite{jagielski2020auditing,nasr2023tight,steinke2023onerun} establish
controlled canaries and statistically valid empirical privacy tests.
Randomized multi-canary audits already include explicit null hypotheses and
decision rules~\cite{pillutla2023randomization}; calibrated membership attacks
likewise emphasise reference distributions and declared operating points
~\cite{carlini2022lira}. We adapt
them to channel-level split-protocol evaluation and add a coverage--amplitude
dose ladder that estimates each metric's onset before the gate is applied.

\subsection{Split-learning attacks, benchmarks, and adjacent systems}

The broader lineage already covers honest-but-curious reconstruction
~\cite{erdogan2022unsplit}, gradient-side label leakage
~\cite{li2022labelleakage}, malicious backward-signal control
~\cite{pasquini2021tiger}, and leakage from intermediate training states
~\cite{gao2023pcat}. SIMBA~\cite{singh2024simba} and VFLAIR-LLM
~\cite{vflairllm} provide modular evaluation substrates; VFLAIR-LLM reports
5 attacks $\times$ 9 defences. Multi-attack split evaluation and bidirectional threat models are therefore
well established. The narrower
contribution here is the measured positive-control onset, channel-specific
shuffled-label negative control, gate set in advance, and artefact-level traceability
in one audited system. Table~\ref{tab:positioning} places this report against
those systems on the axes that separate them.

Within our selected 13-paper audit corpus, five works are in-scope split-LLM
evaluations and eight are adjacent systems (trusted execution environment
partitioning, inference-time obfuscation, on-device key-value cache protection,
prompt sanitisation, and incentive design). MIXGUARD~\cite{mixguard} tunes adaptive
gradient perturbation against
a public proxy and reports defence-vs-attack comparisons, rather than detector
calibration against a shuffled-label negative control. Forward-only inference frameworks
(e.g., \cite{luoprompt,aloepri,gelo}) do not include the training-gradient
channel measured here.

\begin{table}[H]
\centering\scriptsize

\setlength{\tabcolsep}{3pt}
\begin{tabular}{@{}p{0.15\linewidth}p{0.16\linewidth}p{0.13\linewidth}p{0.24\linewidth}p{0.12\linewidth}p{0.11\linewidth}@{}}
\toprule
System & Threat model & Covers training? & Privacy evidence & Integrity & Cost \\
\midrule
\textbf{This work} & compromised cloud; 9 declared channels: 3 measured, 1 constructible, 5 declared unmeasured & yes: 2k-step cells, 6 fresh seeds, 2 datasets; 40k leak cells & forward probe at floor on defended cells (conditional: gate fires $\ge$6\% coverage); joint view breaks the gate on 6/6 cells ($+1.44..2.18$~\pp) until clipping and noise close it (3/3 at floor) & $K{=}3$ median 100\%/0\% (one perturbation class) & $1.05..2.5\times$ WAN \\
AloePri~\cite{aloepri} & honest-but-curious & no (inference) & point TTRSRs; VMA 13.5--25\% on two models & none & $\approx$0\% online; 482 min offline at 671B \\
GELO~\cite{gelo} & VRAM-read $+$ TEE island & no & p95 cosine/Gram error, no control & scoped out & 20--30\% microbenchmark \\
PermLLM~\cite{permllm} & permutation obfuscation & no & broken by the matching family ($>$99\% reconstruction) & none & 3 s/token (6B) \\
ObfNet~\cite{obfnet} & backend holds obfuscators & no & perceptual (10 volunteers) & none & 0.22--11 ms/sample edge \\
TEE~\cite{phalatee} & trust silicon vendor & partial & hardware-rooted attestation & hardware-rooted & $<$7--20\% \\
MPC/HE~\cite{sigma,bumblebee,nexus,puma} & semi-honest parties & no & cryptographic by assumption & by construction & $10^{2}$--$10^{4}\times$ \\
TOPLOC~\cite{toploc} & untrusted provider & no & n/a (integrity only) & hash commits, 100\%/0\% & 258 B / 32 tokens \\
VFLAIR-LLM~\cite{vflairllm} & evaluation harness (5 attacks $\times$ 9 defences) & substrate & shared substrate; its evaluations lack the calibrated null & n/a & harness \\
\bottomrule
\end{tabular}
\caption{Positioning against the closest systems. ``Empirical'' = attacker-measured
privacy; our excess is over a matched no-attack control with a confidence bound
and a pre-declared gate, and the thresholds set in advance are calibrated by injected
leaks (Section~\ref{sec:calibration}). Literature rows report those systems' own published numbers;
their evidence columns use their own authors' metrics, the uncalibrated,
no-matched-null pattern of Section~\ref{sec:related}, so read the comparison as evaluation
\emph{method}, not as relative leak size.}
\label{tab:positioning}
\end{table}

\section{Scope and limitations}
\label{sec:scope}

The findings concern one split-training implementation. The main configuration,
calibration, and shape analyses (Section~\ref{sec:shape}) use WikiText-2; a second corpus is a three-seed
diagnostic robustness check, not evidence of corpus independence. The external
evaluation claim is limited to the targeted comparison of three recent works in
Table~\ref{tab:external}.

Five adversarial families are declared \textbf{unmeasured}, not covered:
membership/property AUC (needs randomised membership assignment), response-side
recovery (no response-side capture), timing metadata (no applicable metric),
stateful remote state, and accumulated history. Active perturbation has an
applicable calibrated metric but is not executed at the scale of the mitigation runs. The
utility side likewise fails on the main configuration ($\Delta$ loss $0.9185$
on the original cell, $0.896$ on the clean rerun, both vs the
0.35 gate). The mitigation runs of
Section~\ref{sec:confirmatory} answer that directly: the gate breaks on six of six cells that pass \emph{both} gates.
Those cells are 2{,}000-step diagnostic runs on the 0.6B model, and
convergence-scale evidence remains out of scope. Rare-token recovery is exactly
zero on the gradient arm under the implemented emitters (the wire arm shows
isolated single-row recoveries, 0.015--0.030\%). TAG-, LAMP-, FILM-, or
DAGER-style sequence reconstruction has not been adapted to this object, a
gradient cut at the split boundary rather than a full parameter
gradient~\cite{deng2021tag,balunovic2022lamp,gupta2022film,petrov2024dager}.
We therefore establish frequent-token discrimination plus exact classification
of the constructed partition, not reconstruction of held-out text.

Additional negative results illustrate the difficulty of the problem. Rotation
alone leaked 18--66\% of tokens across three early rotation-only variants; noise
strong enough to mask the signal destroyed utility first; longer frames leaked
more ($+1.25$ to $+1.36$~\pp{}); and a short private phase after public
pretraining still leaked ($+1.49$~\pp{}). A Mutual Information Neural Estimator
(MINE) reading near zero nats is not a privacy certificate: a finite-sample
lower bound cannot upper-bound true mutual information. The same cell that read
$-1.3\times10^{-6}$ nats held a detected $+0.758$~\pp{} probe effect.

Two bodies of earlier evidence sit outside the argument above and are recorded
in the appendices rather than dropped. Appendix~\ref{app:stack} names each
mechanism of the evaluated defence, its implementation site, and its status.
Appendix~\ref{app:batteries} reports the historical attack batteries against
the defended \emph{forward} cell; they precede the representation-matched
positive control of Section~\ref{sec:positive_control} and are superseded by it
as evidence of probe sensitivity.

Held-out cross-entropy is measured on blocks held out within one flattened
corpus stream rather than on held-out documents (Table~\ref{tab:setup}), so
every held-out decision here (the $0.35$ utility gate, the claim that the
four-layer topology passes it, and the $\approx 0.01$~nat mitigation cost) is a
block-held-out
reading, and claims that depend on document independence are out of scope.

On availability: the protocol, metric thresholds, and family map were fixed
before the replication and mitigation runs, and those fixed versions are the ones in the
artefact release. Committed summaries and code support the displayed
calibration and shape results; raw cluster-side prediction tensors are not part
of the release. A scorer reimplementation, written internally rather than by an
independent group, matched all nine primary seed values to $\le
10^{-6}$~\pp{}; that scorer and its prediction tensors are uncommitted, so the
check is recorded but not re-executable. The complete transcript is likewise
host-only under the raw-data policy. Appendix~\ref{app:artifacts} gives the
result-by-result artefact index, availability boundaries, recorded
verification, and documented limitations.

\section{Conclusion}

This systems-security case study shows a forward-only evaluation passing while
an omitted backward channel revealed which rows were real and which were decoys.
The zero-support construction is an implementation and system-design defect;
the false pass illustrates the more general evaluation failure mode of omitting
an observable channel from the gate. This exact structural metadata disclosure
is distinct from the implemented probe's modest frequent-token paired effect
(claim boundaries are recorded in Section~\ref{sec:scope}). On both datasets, all six defended,
open-gradient mitigation runs passed the forward
privacy and utility gates yet exceeded the privacy gate in the joint view. In
the three mitigated runs, per-row gradient clipping plus Gaussian noise removed
the zero-support partition signal and suppressed the implemented probes below
the gate at a held-out cross-entropy cost of $\approx 0.01$~nats.

Controlled injection, a predeclared per-metric gate, and shuffled-label
negative-control
falsification form the calibrated protocol used to evaluate this case. They are
not evidence for a general methodology across independently designed systems.
In a targeted comparison of three recent evaluations, none combined all three
controls.
The evidence shows that a passing verdict is meaningful only for the channels,
attacks, and leak magnitudes against which its instrument has been tested.

\appendix
\section{Artifact and verification index}
\label{app:artifacts}

All paths in this appendix are relative to the artefact release at
\begin{center}
\artifacturl
\end{center}
The index is organised by claim level rather than by
directory. A committed summary or manifest supports inspection of the displayed
value; ``re-derivable'' means end-to-end execution from that release alone;
availability boundaries are recorded per result below.

\subsection{Headline results}

\subsubsection{Structural metadata disclosure}
The exact real/decoy split is recorded for each seed under
\path|paper-data/collected/diagnostic/e1_reproduction_w12/|. The primary
per-seed bundle summary is named
\path|e1_repro_w12_s44_bundles.json| for seed 44, with corresponding files for
the other seeds. These committed derived records support the frame and row
agreement values; the underlying cluster-side prediction tensors are not
committed. The complete verified transcript covers 3 seeds, 30,000 optimiser
steps, and 180,636 events in the committed manifest. The hardened verifier and
the forward, gradient, and joint-view consumers confirm that all 30,000
training frames carry all four payload directions. Payload-level attacks on
the accumulated history remain unexecuted, and the raw transcript is
host-only. The committed transcript index is under
\path|paper-data/collected/diagnostic/w34_complete/|.

\subsubsection{Frequent-token content inference}
The exploratory per-seed paired summaries and shuffled-label negative controls
are stored with the structural records above. The twelve-seed hierarchical
estimate is at
\path|paper-data/collected/diagnostic/e1_hierarchical/| and is generated by
\path|bin/summarize_complete_view_matrix.py|. The representation-matched
positive-control records are under
\path|paper-data/collected/diagnostic/deep_probe/|; the distinct
isolation-audit rerun is under
\path|outputs/deep_probe_validation_2026-08-27/|.

\paragraph{Seed-44 verification trace.}\mbox{}\par
The committed per-seed paired record is stored under

\noindent\path|paper-data/collected/diagnostic/e1_reproduction_w12/|.
The record is named \path|e1_repro_w12_s44_arm_grad_real_paired.json|. Read
\texttt{best\_eligible.paired\_advantage\_pp}: the value $+0.6929$
reproduces Table~\ref{tab:sixseed} row-wise. The paired statistic behind that
field is computed by \path|bin/paired_advantage.py| against the constant
baseline over frame-clustered evaluation rows. The adjacent negative-control
record,
\path|e1_repro_w12_s44_arm_grad_real_shuffled_paired.json|, must read at floor.
This is the recorded verification trace for the reported value.

\subsubsection{Calibration and attack-specific shape}
The declared-channel table uses \path|paper-data/family_metric_map.json|
together with the mitigation-run results, and is regenerated with
\path|bin/build_channel_table.py|. The dose--response source is
\path|paper-data/collected/diagnostic/w24_metric_sweep/w24_dose_response.json|,
the associated summaries are under
\path|paper-data/collected/diagnostic/w24_metric_sweep/|, and the figure command
is \path|papers/paper-1/figs/build_figures.py|. The threshold map is
\path|paper-data/family_metric_map.json|. The nine-cell shape records comprise
the \path|paper-data/collected/diagnostic/gradaudit/| cell JSONs. Their full
summary is

\noindent{\small\path|paper-data/collected/diagnostic/gradaudit/w56_gradaudit_summary.json|}.

They use the same \path|papers/paper-1/figs/build_figures.py| command. These
calibration and shape rows are re-derivable from committed summaries and code.

\subsection{Confirmation}

\subsubsection{The replication runs}
Seeds 48--50 are retained under
\path|paper-data/collected/diagnostic/e1_confirmation/|. Their committed
paired summaries support the displayed values; raw prediction tensors remain
cluster-side.

\subsubsection{The mitigation runs}
The mitigation-run cells, paired statistics, and SHA-256 manifest of the 7.4\,GB bundle
store are committed under
\path|paper-data/collected/diagnostic/phasec_2026-08-27/|. Per-cell records are
named \path|paired_summary.json|. The packaged driver and scorer redisplay the
committed derived summaries; the raw bundle store remains on the cluster.

\subsubsection{Internal scorer availability}
The independent scorer check is recorded in
the internal red-team review of the scorer reimplementation
(\path|docs/audits/W54_RED_TEAM_2026-08-26.md|). It re-derived all nine primary
seed values to $\le 10^{-6}$~\pp{}. The report is committed, but the independent
scorer and its recorded prediction tensors are not, so this verification is
recorded but not repository-re-executable.

Table~\ref{tab:availability} consolidates the availability boundary for these
result categories.

\begin{table}[H]
\centering\scriptsize
\begin{tabularx}{\textwidth}{@{}>{\raggedright\arraybackslash}p{2.7cm}>{\raggedright\arraybackslash}X>{\raggedright\arraybackslash}p{3.5cm}>{\raggedright\arraybackslash}p{2.5cm}@{}}
\toprule
result category & committed input & committed code & re-derivable \\
\midrule
exploratory gradient result & per-seed paired JSONs and negative controls; bundles are manifests to cluster-side prediction tensors & packaged scorer & displayed values from committed summaries \\
nine audit cells & \path|paper-data/collected/diagnostic/gradaudit/| cell JSONs and summary & \path|papers/paper-1/figs/build_figures.py| & yes \\
dose--response calibration & \path|paper-data/collected/diagnostic/w24_metric_sweep/| JSONs & \path|papers/paper-1/figs/build_figures.py| & yes \\
mitigation runs & \texttt{phasec\_2026-08-27/} manifests and paired summaries & packaged driver and scorer & displayed values from committed summaries \\
internal scorer check & internal-check report & uncommitted independent scorer & recorded verification only \\
complete transcript & hashes and manifest & verifier & no; host-only by policy \\
\bottomrule
\end{tabularx}

\caption{Per-result availability. The audit-cell and calibration rows
regenerate from committed summaries; the exploratory and mitigation rows are
redisplayed from committed derived artefacts whose raw prediction tensors
remain on the cluster. Transcript hashes and the manifest are committed, but
the payload remains on its host.}
\label{tab:availability}
\end{table}

\subsection{Corrections and limitations}

\subsubsection{Sequential-block split versus the document protocol}
The \path|paper-data/evaluation_protocol.json| file, fixed in advance, declares a document-level held-out
split, but the committed runner implements the sequential fixed-width block
split reported in Table~\ref{tab:setup}. Held-out measurements are therefore
block-held-out within one flattened corpus stream, not document-held-out;
claims that depend on document independence remain out of scope. This records
the discrepancy; the protocol file itself was left unedited.

\subsubsection{Corrections of record}
The traceability verification is recorded at
the traceability audit, which fixes the committed-versus-cluster-side evidence
boundary (\path|docs/audits/W78_TRACEABILITY_2026-08-26.md|). It supersedes earlier
wording about what is committed versus cluster-side. The adversarial review
round, an external adversarial re-read of the manuscript's evidence claims,
at \path|docs/audits/W78_CODEX_REVIEW_2026-08-27.md| applies it, and the
same red-team report records
the independent re-derivation. The traceability audit has precedence for
evidence boundaries.

A forward-membership reading of $+0.068$ (AUC above the 0.5 floor, stable
across all six packaged seeds) was falsified by the shuffled-label negative
control: it survives label permutation and randomised membership assignment,
decomposes into in-sample memorisation ($+0.095$) with a corpus-region term of
$\approx 0$, and exists only in the coordinate probe that cannot generalise.
The purported members were the probe's own training rows. We withdrew the channel; and \texttt{membership\_auc} remains a diagnostic. The underlying capture remains on the cluster.

Claims withdrawn during the study are recorded in
\path|paper-data/claim_evidence_ledger.json| with their refutation records; the
appendices that follow retain withdrawn entries only to identify their status
and provenance.

\section{The defended stack, mechanism by mechanism}
\label{app:stack}

Table~\ref{tab:stack} names each mechanism of the evaluated defence concretely,
with its implementation site and its status in the current evidence. Three rows
carry a non-core status, each for a recorded reason: the fragmentation cell's
remote modules never received a gradient (a dispatch defect; the cell is
invalidated as a trained-fragmentation experiment), the public-pretraining and
capacity framing are withdrawn per the refutation record, and the deep defended
cloud's capacity conclusion is withdrawn with it.

\begin{table}[H]
\centering\scriptsize

\setlength{\tabcolsep}{3pt}
\begin{tabular}{@{}p{0.12\linewidth}p{0.24\linewidth}p{0.48\linewidth}p{0.09\linewidth}@{}}
\toprule
Purpose & Technique & Detail & Status \\
\midrule
Compression & adversarially trained MLP encoder/decoder (information bottleneck) & $H{\to}D{=}64$ MLP with GELU, trained by minimax against embedded token/property/reconstruction probes plus a MINE MI penalty (the MI reading is not a certificate) & core \\
Noise & per-token-row L2 clip to $C{=}1.0$ $+$ additive Gaussian $\sigma{=}0.35{\cdot}C$ (0.40 on hybrid MoE), both directions & Gaussian mechanism with zCDP accounting (the $\varepsilon$ is quoted as vacuous) & core \\
Key derivation & SHA-256 counter-mode KDF & 128-bit CSPRNG master per draw, domain-separated \texttt{sha256(DOM$\|$S$\|$t)}, exact $(u64{\gg}11){\cdot}2^{-53}$ uniforms, float64 Box-Muller & core \\
Rotation & Haar-random orthogonal & Gaussian matrix $\to$ QR with positive-diagonal sign fix, at $D$ width; fresh per request & core \\
Permutation & Fisher--Yates over the KDF stream & without replacement; fresh per request & core \\
Scale gauge & per-row $\exp(N(0,\sigma))\cdot(\pm1)$, $\sigma{=}0.75$, clamped $[0.2,5]$ & independent of the other gauges (separate KDF epoch); OFF for radial/invariant-MLP clouds & core \\
Decoy rows & recycled \emph{real} latent rows from earlier train blocks & CSPRNG-sampled without replacement, honest labels tracked; decoys are real data, not noise & core \\
Transport & websockets over TLS~1.3 & pinned self-signed CA (trusts no system store), hostname verification on, plaintext refused & core \\
Session isolation & fresh cloud model + AdamW per connection & seeded per session under \texttt{fork\_rng}; no cross-session state & core \\
Gradient hardening & per-token L2 clip 1.0, non-finite $\Rightarrow$ abort & on remote-returned gradients & core \\
Integrity & identical-seed replicas $+$ median compare & relative deviation from group mean $>$ 0.02 $\Rightarrow$ flagged; median neutralises one adversary; demonstrated on one large perturbation class only & avail. \\
Cloud compute & gauge-equivariant Gram message passing & softmax over squared unit-Gram $\times$ Gram @ unit rows (monomial), expert routing, or invariant-feature MLP gates & core \\
Wire quantisation & fixed-grid int8 over the gauge clamp range & straight-through on the defender gradient path; absmax variant rejected (strips the scale gauge) & avail. \\
Replay resistance & per-request nonce acceptance window & a replayed frame is refused & core \\
Key ratchet & per-epoch key derivation on the KDF chain & limits cross-request capture pooling; the current transform is necessarily visible to the executor; it bounds accumulation, it is not irreversibility & core \\
Redundant channels & $K{=}2$ independent cloud channels, fresh gauges per channel & cross-check + independence & core \\
Two-segment delegation & cloud segment A / private island / cloud segment B & chained surrogate error compounds across segments, the measured reason it stays optional & opt. \\
Public pretraining & UCN trains on public data at full fidelity; private phase re-hardens & the short-private-phase cell leaked ($+1.49$~\pp{}); privacy cost zero on the public stream & opt. \\
Fragmentation & frame split into two $D{=}32$ half-views & structural partition across servers; the measured cell's remote modules never trained, invalidating it as a trained-fragmentation experiment & invalid. \\
Gram flattening & regularizer penalising off-diagonal Gram concentration & keeps the invariant features uninformative & avail. \\
Deep defended cloud (isolation cells) & deep invariant-feature MLP: norms and Gram-statistic feature channel with deep wide gates & 118.9M params; its capacity conclusion is withdrawn & withdr. \\
Complementary views & $K{=}3$ disjoint 64-wide views, one per cloud, per-view rotation, one shared row permutation & privacy rests on k-of-K collusion; union attacker evaluated conservatively & opt. \\
Secret mixup & released rows $= \lambda z + (1{-}\lambda) z_{\text{decoy}}$, $\lambda{=}0.75$ & TLN corrects the return via a parallel decoy forward; approximate under the nonlinear cloud, cost measured & opt. \\
Adaptive controller & fresh probe every 100 steps; tripwire actions + utility-gate width ladder & one gross breach repaired within one window; its monitor was later shown blind at relevant operating points & withdr. \\
\bottomrule
\end{tabular}
\caption{Techniques at a glance: what each mechanism \emph{is}, concretely,
where it lives, and its status (core / available / optional / withdrawn /
invalidated) in the current evidence.}
\label{tab:stack}
\end{table}

None of these are novel primitives in isolation. QR rotations, Fisher--Yates,
Gaussian mechanisms, TLS are all standard. The engineering content is their
composition order and their measured interaction: the bottleneck alone leaks,
the gauges alone leak, noise alone destroys utility. The composite at these
settings is what sits at the floor. That is why the table names not just the
technique but the configuration it was validated at.

\section{The defended cell under its historical attack batteries}
\label{app:batteries}

The three batteries below attack the \emph{defended forward} cell (the
conditional claim), not the backward channel this report is about. Each uses
one probe family on the forward view only, and the battery's convention reads
\emph{trends} below a weak constant baseline. They are reported here with the
representation-matched positive control
(Section~\ref{sec:positive_control}, Table~\ref{tab:positive_control}) that the
original batteries lacked.

These batteries were run on two models: the Qwen3-0.6B of
Table~\ref{tab:setup}, and a 35B-A3B mixture-of-experts model used only here,
as a capacity check on the defended forward cell. No result in the body of this
report depends on the larger model. Table~\ref{tab:battery} reports the
compromise-fraction battery, Table~\ref{tab:matching} the external
matching-attack families, and Table~\ref{tab:byz} the Byzantine median
verification.

\begin{table}[H]
\centering\scriptsize

\begin{tabular}{@{}p{0.27\linewidth}p{0.20\linewidth}p{0.15\linewidth}p{0.30\linewidth}@{}}
\toprule
Arm (fraction grid) & 0.6B trend & 35B-A3B trend & Reading \\
\midrule
Wire capture 10/25/50/100\% & flat, $-4.7$ to $-4.9$ & flat, $-4.6$ to $-5.1$ & capturing the whole wire does not help \\
Decoy identities revealed 0/25/50/90\% & $-4.9$ to $-7.4$ & $-4.9$ to $-5.6$ & knowing which rows are decoys makes the probe worse \\
Gauge secrets leaked, 10--20\% & $-7.8$ ungauged / $-4.7$ gauged & $-5.7$ / $-4.6$ & rows seen before gauging still read at guessing level \\
Known plaintext 1/5/20\% & flat, $-3.8$ to $-4.4$ & flat, $\approx-4.0$ & knowing some tokens does not reveal the others \\
\bottomrule
\end{tabular}
\caption{Compromise-fraction battery on the defended cell. Each row gives the
attacker a growing share of some secret and asks whether the probe improves.
Values are the probe's accuracy in percentage points above or below its
constant-guess baseline, so a negative number means the probe did worse than
guessing. None of these arms approaches the $+1.0$~\pp{} gate, and none
improves as the attacker is given more.}
\label{tab:battery}
\end{table}

\begin{table}[H]
\centering\scriptsize

\begin{tabular}{@{}lrrr@{}}
\toprule
Model & Gate statistic (pp) & Vector-matching top-1, against baseline & Nearest-signature top-1, against baseline \\
\midrule
0.6B (seed 42) & $+0.541$ & $0.03\%$ vs $5.92\%$ & $1.56\%$ vs $5.92\%$ \\
35B-A3B (seed 42) & $+0.285$ & $0.02\%$ vs $5.85\%$ & $1.54\%$ vs $5.85\%$ \\
\bottomrule
\end{tabular}
\caption{External matching-attack families vs.\ the defended cell (regenerated
bundles). Two adapted attack families on these captures. The positive-control
sensitivity they presuppose is Table~\ref{tab:positive_control}.}
\label{tab:matching}
\end{table}

\begin{table}[H]
\centering\scriptsize

\begin{tabular}{@{}lrrrrr@{}}
\toprule
Cell & Verified frames & Flagged & Max rel.\ dev. & Loss $\Delta$ & Ratio \\
\midrule
3 honest CPU nodes & 2{,}256 & 0 (0\% FP) & $4.98{\times}10^{-8}$ & $+0.071$ & $9.30\times$ \\
2 honest $+$ 1 malicious (lying) & 2{,}256 & 2{,}256 (100\%) & $0.5947$ & $+0.088$ & $9.13\times$ \\
\bottomrule
\end{tabular}
\caption{$K{=}3$ Byzantine median verification. One large perturbation
class on identical CPU replicas; no general active-security claim.}
\label{tab:byz}
\end{table}

\FloatBarrier
\bibliographystyle{plain}
\bibliography{refs}

\end{document}